\documentclass[preprint,showpacs,preprintnumbers,amsmath,amssymb, aps]{revtex4}

\usepackage{color,graphicx}% Include figure files
\usepackage{dcolumn}% Align table columns on decimal point
\usepackage{bm}% bold math

\begin{document}

\preprint{Preprint Copy}

\title{Quenched magnetic moment in Mn-doped amorphous Si (\textit{a}-Mn$_{x}$Si$_{1-x}$) across the metal-insulator transition}% Force line breaks with \\

\author{Li Zeng}
\email{lzeng@ucsd.edu}
  \affiliation{Materials Science and Engineering Program, University of California, San Diego, La Jolla, CA, 92093, USA}

\author{E. Helgren}
\author{F. Hellman}
 \affiliation{Department of Physics, University of California, Berkeley, Berkeley, CA, 94720, USA} 

\author{R. Islam}
\author{B. J. Wilkens}
\affiliation{Center for Solid State Science, Arizona State University, Tempe, AZ, 85287, USA}

\author{R. J. Culbertson}
\author{David J. Smith}
 \affiliation{Department of Physics, Arizona State University, Tempe, AZ, 85287, USA}
%\homepage{}

\date{\today}% It is always \today, today,

\begin{abstract}
The magnetic and electrical transport properties of Mn-doped 
amorphous silicon (\textit{a-}Mn$_{x}$Si$_{1-x}$) thin films 
have been measured. The magnetic susceptibility obeys the 
Curie-Weiss law for a wide range of $x$ (0.005-0.175) and the 
saturation moment is small. While all Mn atoms contribute to 
the electrical transport, only a small fraction (interstitial 
Mn$^{2+}$ states with $J$=$S$=5/2) contribute to the magnetization. 
The majority of the Mn atoms do not possess any magnetic moment, 
contrary to what is predicted by the Ludwig-Woodbury model for 
Mn in crystalline silicon. Unlike \textit{a-}Gd$_{x}$Si$_{1-x}$ 
films which have an enormous \textit{negative} magnetoresistance, 
\textit{a-}Mn$_{x}$Si$_{1-x}$ films have only a small 
\textit{positive} magnetoresistance, which can be understood by 
this quenching of the Mn moment.
\end{abstract}

                             % Classification Scheme

%\keywords{amorphous silicon, magnetic semiconductor}%Use showkeys class option if keyword
\maketitle

\section{\label{sec:level1}introduction}
%\subsection{\label{sec:level2}}
%\subsubsection{\label{sec:level3}}

Mn has been widely used as a magnetic dopant in dilute magnetic semiconductors (DMS), which have potential applications as spintronic materials. While many Mn-doped systems are antiferromagnetic (AFM), ferromagnetism was found in Mn-doped III-V semiconductors, such as GaMnAs, first reported by Ohno~\cite{Ohno1998Science}. Both mean-field theory (Zener model)~\cite{Dietl2000} and first-principles calculations~\cite{Stroppa2003, Bernardini2004, Weng2005} have predicted that room temperature ferromagnetism is possible in Mn-doped group-IV semiconductors. An empirical rule is that short range Mn-Mn direct interaction favors AFM coupling, while long-range indirect interactions (RKKY, double exchange, kinetic exchange) favors ferromagnetic ordering. Interactions and mixing of the Mn \textit{d} electrons in different host environments will lead to different charge and spin states, which greatly affect the magnetic ordering~\cite{Weber1983}. There is a strong recent focus on research into Mn-doped crystalline Ge (\textit{c}-Ge) and Si (\textit{c}-Si)~\cite{Park2002, Kang2005PRL, Kwon2005, Bolduc2005, Awo2006, Zhou2007PRB}. Mn-doped \textit{c}-Si, in particular, is of much interest due to its compatibility with Si-based semiconductor technology. However, the solubility of Mn in crystalline group-IV semiconductors is extremely low. Therefore, to stabilize Mn doping above the solid solubility limit, one must utilize highly non-equilibrium doping techniques such as low temperature molecular beam epitaxy (MBE)~\cite{Zhang2005JAP} or ion-implantation~\cite{Weber1983, Park2002}. Recent studies on Mn-implanted crystalline Si (\textit{c-}Si$_{1-x}$:Mn$_{x}$) thin films prepared by the latter technique have achieved Mn doping levels of a few atomic percent ($x$$\leq$0.05) ~\cite{Kwon2005, Bolduc2005}. 

Ferromagnetism [Curie temperature ($T_{C}$) up to 116 K] was observed in Mn-doped \textit{c-}Ge (\textit{c-}Mn$_{x}$Ge$_{1-x}$) and a long-range ferromagnetic (FM) interaction dominating over a short-range Mn-Mn AFM interaction was proposed as the origin of the ferromagnetism~\cite{Park2002}. Determining whether the ferromagnetism in Mn-doped \textit{c}-Ge and \textit{c}-Si is intrinsic or not in origin, is however, very challenging due to thermodynamically favorable FM second phases or possible nano-structures, which are hard to detect by conventional structural characterization tools. Recent studies, including sub-micron scanning photoelectron microscopy (SPEM), high-resolution transmission electron microscopy (HRTEM) and synchrotron x-ray diffraction, have shown chemical inhomogeneity in \textit{c}-Ge$_{1-x}$Mn$_{x}$~\cite{Kang2005PRL} and the existence of MnSi$_{1.7}$ nano-crystallites in \textit{c-}Si$_{1-x}$:Mn$_{x}$~\cite{Zhou2007PRB}. The ferromagnetism in these systems was attributed to a Mn-rich Ge phase~\cite{Kang2005PRL} and MnSi$_{1.7}$ nano-crystallites~\cite{Zhou2007PRB} in the former and latter systems, respectively.

One way to stabilize homogeneous Mn doping is to prepare the film in the amorphous form. A strong attraction of a high quality amorphous matrix is its outstanding flexibility to incorporate high concentration of dopant atoms with large radius, such as Gd~\cite{Zeng2007PRB}. For up to 19 at.\% Gd doping, the structure of the Gd-Si solid solution is arranged such that its atomic number density is almost the same as that of pure amorphous Si (\textit{a}-Si). While the amorphous structure may alter the long-range magnetic ordering (especially for AFM materials), the localized magnetic moment at each doping site should still be preserved since charge and spin states are primarily determined by the local environment. Instead of directly addressing the question of whether the final solid-state is ferromagnetic, antiferromagnetic or paramagnetic, it is conceptually easier to understand the existence of local magnetic states, which are the building blocks for more complicated magnetic interactions and orderings. This conceptual simplification was suggested to be of great value in understanding the ferro- and antiferro-magnetism in transition-metal solid solutions~\cite{Anderson1961}, and may also apply to the study of transition-metal-doped semiconductor alloys. 

Metallic dopants in amorphous semiconductors can also provide charge carriers from delocalized $d$ and $s$ electrons, which contribute to the electrical conductivity. A concentration-tuned metal-insulator (\textit{M-I}) quantum phase transition has been reported in a variety of doped amorphous semiconductors, such as \textit{a}-Nb$_{x}$Si$_{1-x}$~\cite{Hertel1983}, \textit{a}-Gd$_{x}$Si$_{1-x}$~\cite{Teizer2000SSC, Teizer2000PRL} and \textit{a}-Gd$_{x}$Ge$_{1-x}$~\cite{Sinenian2006GdGe}. In \textit{M-I} physics, a metal is defined as having finite dc conductivity ($\sigma_\text{dc}$) as $T$$\rightarrow$0 K, whereas an insulator is defined by $\sigma_\text{dc}$$\rightarrow$0 when $T$$\rightarrow$0 K. If the dopant has a large local moment, such as Gd$^{3+}$ with $J$=$S$=7/2, thin film samples will have an enormous negative magnetoresistance (MR$\sim$10$^{5}$ at 1 K), which leads to a magnetic-field-induced \textit{M-I} transition in \textit{a}-Gd-Si and \textit{a}-Gd-Ge alloys~\cite{Teizer2000PRL, Sinenian2006GdGe}. The mechanism for this enormous negative MR is not fully understood, but is clearly related to a carrier-moment exchange interaction and magnetically-induced disorder in zero magnetic field which leads to carrier localization. Electron correlation effects and electron screening have been suggested to play an important role~\cite{Helgren2005T*} in the magnitude of the MR and its onset temperature, denoted as $T^{*}$. 

There have been limited reports about Mn-implanted \textit{a}-Si (denoted as \textit{a}-Si$_{1-x}$:Mn$_{x}$), mainly focusing on transport properties for $x$$\sim$0.07-0.22~\cite{Yakimov1995, Yakimov1997_JPhys, Yakimov1997_JETP}. Samples were prepared by room-temperature Mn ion-implantation either into $e$-beam-evaporated \textit{a-}Si on quartz ($x$$\leq$0.13)~\cite{Yakimov1995} or into \textit{c-}Si on sapphire with simultaneous amorphization of the films ($x$$>$0.13)~\cite{Yakimov1997_JPhys}. No magnetic properties were reported. However, a characteristic temperature was observed in the transport data, which was attributed to spin-glass (SG) ordering by $d$-$d$ spin correlation between Mn atoms. The characteristic SG temperatures ($T_\text{SG}$) were not confirmed by magnetization measurements, but were estimated from the cross-over temperature between a variable-range hopping (VRH) behavior and a simple-activated behavior obtained from dc conductivity data fitting. From this analysis, $T_\text{SG}$ was claimed to vary from 6 to 20 K, depending on the Mn concentration. This $T_\text{SG}$ is rather high compared to that found in amorphous MnSi ($x$=0.5), with SG freezing at $T_{f}$=22 K. Such high $T_\text{SG}$ requires a local Mn moment with $d$-$d$ exchange interactions. These results were explained by AFM ordering of small magnetic clusters and magnetic polaron formation within the Mn clusters. The \textit{a}-Si$_{1-x}$:Mn$_{x}$ films undergo an \textit{M-I} transition at a similar critical composition ($x_{c}$$\sim$0.137~\cite{Yakimov1997_JETP}) as for \textit{a}-Gd$_{x}$Si$_{1-x}$, but there is no large MR effect~\cite{Yakimov1995}; the obvious difference is \textit{d} vs \textit{f} electron magnetism. The MR must be related to an exchange interaction between carriers and moments. Therefore, it is puzzling that MR would be smaller in a material with $d$ electrons (\textit{a-}Mn-Si) than one with $f$ electrons (\textit{a-}Gd-Si) since $s$-$d$ exchange is typically much larger than $s$-$f$ exchange. 

Ludwig and Woodbury have developed a model to predict the spin and 
charge states of 3$d$ transition metals within 
\textit{c}-Si~\cite{Ludwig1960}. The Ludwig-Woodbury (LW) model 
states that crystal-field splitting is small in \textit{c-}Si 
compared to the on-site exchange energy, so that all 3$d$ impurities 
have a maximum spin value (Hund's rule). The LW model 
was later challenged for the early and late transition metals such as Ni and Ti, 
but is still believed to account for Mn impurities. 
Electron-spin resonance (ESR) experiments confirmed the spin states of 
four different charge states for interstitial Mn (Mn$_\text{I}^{-}$, Mn$_\text{I}^{0}$, 
Mn$_\text{I}^{+}$ and Mn$_\text{I}^{2+}$)~\cite{Weber1983}, all consistent with the LW model. 
The 4 nearest neighbors (tetrahedral symmetry) 
and the 6 second nearest neighbors (octahedral symmetry) of an interstitial site 
in \textit{c}-Si are very close together and thus both sets of neighbors take part in determining the crystal-field splitting. 
This splits the $d$ levels into $e$ and $t_{2}$ levels, 
with $e$ levels lying higher than $t_{2}$ levels. Spin states of the two possible substitutional states 
(Mn$_\text{Si}^{+}$ and Mn$_\text{Si}^{2-}$) proposed by the LW model have not been observed by ESR measurements, but would have $J$=$S$=1 and $J$=$S$=5/2, respectively. 
The $t_{2}$ levels lie higher than the $e$ levels for both Mn$_\text{Si}$ states. 
In \textit{a}-Si, the site symmetry is lowered due to disorder. 
The effect on the crystal field splitting due to the disorder is unknown. 

We report here on Mn-doped amorphous silicon (\textit{a-}Mn$_{x}$-Si$_{1-x}$) thin films, with $x$ between 0.005 to 0.175, covering both the dilute and the \textit{M-I} transition regions. The films were prepared by $e$-beam co-evaporation of Mn and Si under ultrahigh vacuum (UHV) conditions. Co-deposited \textit{a}-Mn$_{x}$Si$_{1-x}$ samples have a much more homogeneous Mn distribution in the \textit{a}-Si matrix compared to that of implanted samples, and should thus provide a better representation of the behavior of this type of amorphous solid solution. The focus of this current work is to understand the magnetic and magneto-transport properties of Mn dopants in \textit{a}-Si, especially the presence or absence of any local Mn moment and its effect on electrical transport. 

\section{experimental details}
The \textit{a}-Mn$_{x}$Si$_{1-x}$ samples were grown by $e$-beam co-evaporation of high purity Si and Mn sources onto substrates held near room temperature (below 60 $^{\circ}$C during film growth). This condition is in contrast to the film preparation described in Refs.~\onlinecite{Bolduc2005} and~\onlinecite{Zhou2007PRB}, where the substrates were held at $\sim$350 $^{\circ}$C to avoid amorphization, and post-implantation annealing (up to 800 $^{\circ}$C) was used to create \textit{c-}Si$_{1-x}$:Mn$_{x}$ films. Our base pressure prior to deposition was below 8$\times$10$^{-10}$ Torr. Thickness monitors were used for each source during deposition in order to precisely control the real-time Mn and Si flux to achieve uniform doping profiles as well as the desired Mn concentrations. The film compositions and lack of oxygen impurities were determined by Rutherford backscattering (RBS), using oxygen-resonance energy to enhance sensitivity, and by high-resolution cross-sectional transmission electron microscopy (HR-XTEM). Magnetic and magneto-transport measurements were made with a superconducting quantum interference device (SQUID) magnetometer from Quantum Design. Further details can be found in Ref.~\onlinecite{Zeng2007PRB}.  

\section{Experimental Results}
\subsection{Material and Structural Characterization}
\begin{figure}
    \centering
    \includegraphics[width=0.4\textwidth]{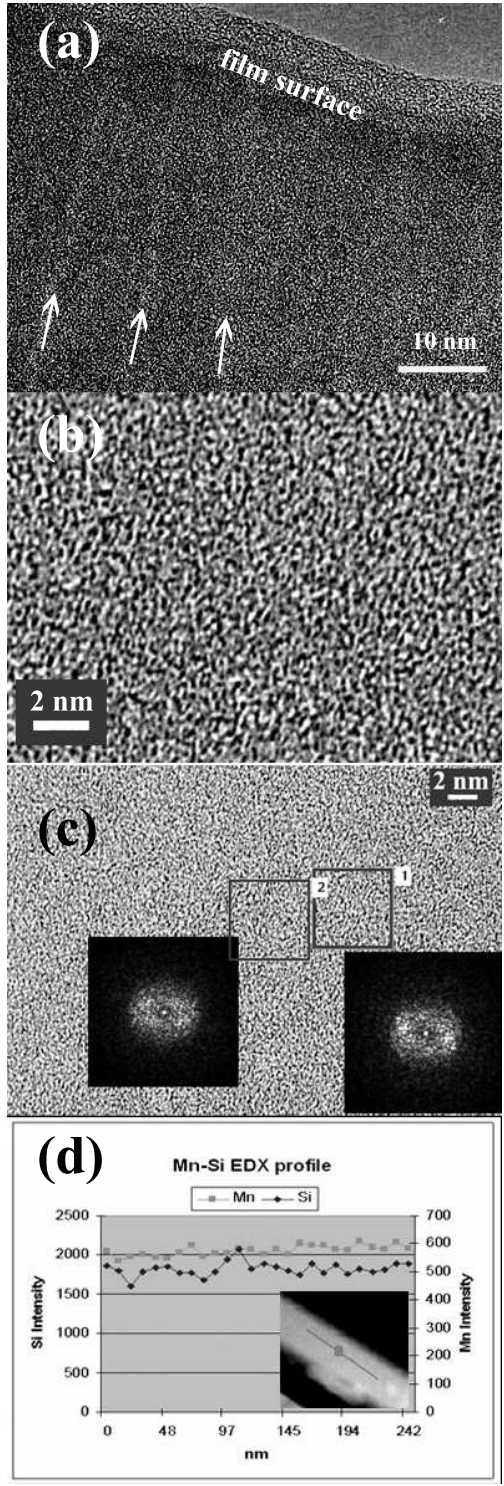}% Here is how to import EPS art
            \caption{(a) Low-resolution XTEM micrograph; arrows specify columnar structure and growth direction;  (b) HR-XTEM micrograph, (c) diffractogram data and (d) EDX profile for \textit{a}-Mn$_{0.11}$Si$_{0.89}$. The scale bars are 10, 2, 2 nm in (a), (b) and (c) respectively.}
            \label{fig:HRXTEM}
\end{figure}

Figure~\ref{fig:HRXTEM} shows HR-XTEM analysis for a typical \textit{a}-Mn$_{x}$Si$_{1-x}$ film, where $x$=0.11. The film shows columnar morphology in the growth direction, a typical feature for low $T$ thermally evaporated thin films~\cite{Zeng2007PRB}, with column diameters $\sim$10 nm [Fig.~\ref{fig:HRXTEM}(a)]. The physical transport and magnetic properties were shown experimentally to not depend on microstructure at this length scale in amorphous Gd-Si alloys prepared by the same growth technique~\cite{Zeng2007PRB}. The \textit{a}-Mn$_{x}$Si$_{1-x}$ films appear to be amorphous with no significant sign of phase segregation or nano-crystallites. Figure ~\ref{fig:HRXTEM}(b) shows a slight suggestion of lattice fringes for regions of less than 2 nm; but digital diffractograms show typical amorphous rings, indicating absence of long-range ordering [Fig.~\ref{fig:HRXTEM}(c)]. Annular-dark-field (ADF) images with energy-dispersive X-ray (EDX) spectroscopy were also used to probe the homogeneity of the \textit{a-}Mn$_{x}$Si$_{1-x}$ films. As an example, Fig.~\ref{fig:HRXTEM}(d) shows the variation of Mn and Si counts along a 250 nm line parallel to the film surface. The Mn distribution was uniform along the scanning length, with no strong indication of significant fluctuation in Mn concentration. 

\begin{figure}
    \centering
    \includegraphics[width=0.5\textwidth]{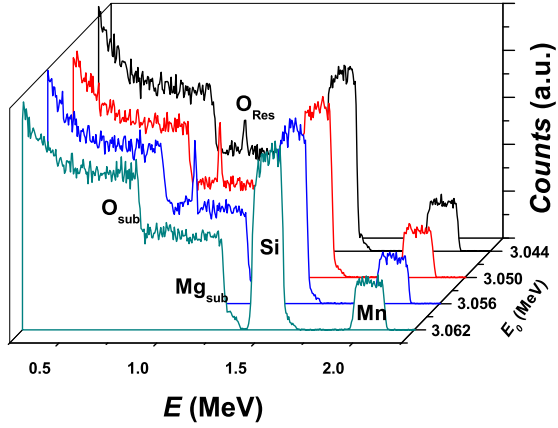}% Here is how to import EPS art
            \caption{(color online) O resonance RBS data for \textit{a}-Mn$_{0.08}$Si$_{0.92}$. All other samples show the same trends. Mg$_\text{sub}$ and O$_\text{sub}$ are signals from MgO substrate. O$_\text{Res}$ is the O resonance signal.}
            \label{fig:RBS}
\end{figure}

Figure~\ref{fig:RBS} shows RBS spectra for a typical \textit{a}-Mn$_{x}$Si$_{1-x}$ film ($x$=0.08). Film compositions were obtained from data simulations with error bars of $\sim$ 0.001 and 0.005, for Si and Mn respectively. Oxygen-resonance spectroscopy was used to probe specifically for oxygen contamination (the oxidation enthalpy of Mn is high compared to Si, and would counteract Mn doping and alter the Mn magnetic properties). A He$^{2+}$ ion beam with O resonance energy ($E_\text{Res}$=3.05 MeV) was used and oxygen-resonance RBS peaks (labeled as O$_\text{Res}$ in Fig.~\ref{fig:RBS}) were observed at $\sim$1.1 MeV, indicating a surface oxide layer. By slightly increasing $E_{0}$ above $E_\text{Res}$, oxygen content below the film surface could be probed (the oxygen resonance occurs when the He$^{2+}$ kinetic energy is reduced to 3.05 MeV after penetrating to a depth below the film surface). If O$_\text{Res}$ were observed over a range of $E_{0}$$>$$E_\text{Res}$, it would indicate oxygen contamination throughout the film. Here, however, the O$_\text{Res}$ peak intensity is reduced and eventually disappears after $E_{0}$ is increased above $E_\text{Res}$, as shown in Fig.~\ref{fig:RBS}. We can therefore conclude that O is only located near the film surface, within less than a few nm, which is small compared to the total film thickness ($\sim$400 nm).

The room-temperature (RT) conductivity ($\sigma_\text{RT}$) of the films as a function of time ($t$) after exposure to air from vacuum was monitored and found to be stable up to at least 200 hours. The slightly reduced $\sigma_\text{RT}$ ($\leq$4\%) for this time period is probably due to a thin self-limiting oxide layer, which is insulating and thus reduces the effective thickness of the film. For this 4\% reduction, the estimated thickness of this oxide layer is less than 10 monolayers, in good agreement with the RBS O resonance data reported above. 

Figure~\ref{fig:Ndensity} shows the total atomic number density ($n_\text{total}$) for different \textit{a}-Mn$_{x}$Si$_{1-x}$ samples, obtained by dividing the RBS areal density by the film thickness, as measured by atomic force microscopy. Since the structure of high quality \textit{a}-Si is a continuous random network (CRN), which preserves the local $sp^{3}$ tetrahedral covalent bonding, with small deviations in bond length and bond angle, the $n$ for \textit{a}-Si should still be comparable to that of \textit{c}-Si~\cite{Zeng2007PRB}. Undoped \textit{a}-Si with 98\% number density of \textit{c}-Si is achieved by $e$-beam evaporation~\cite{brodsky1972}. In Fig.~\ref{fig:Ndensity}, the $n_\text{total}$ of \textit{a}-Mn$_{x}$Si$_{1-x}$ increases with $x$, as opposed to results in the literature for crystalline \textit{c}-X$_{1-x}$:Mn$_{x}$ (X$=$Si, Ge), where additional Mn was found to decrease  $n_\text{total}$ (i.e. increase the lattice constant) because of lattice strain caused by the larger radius of Mn atoms~\cite{Cho2002PRB, Kwon2005, Ma2006}. 

\subsection{Conductivity and Magnetization}
\begin{figure}
    \centering
    \includegraphics[width=0.5\textwidth]{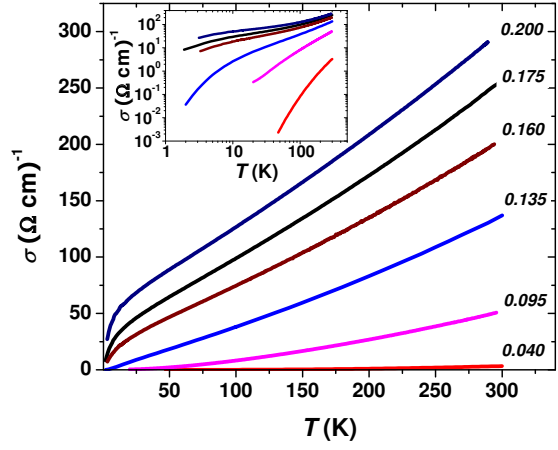} % Here is how to import EPS art
            \caption{(color online) $\sigma_\text{dc}$ vs. $T$ for \textit{a}-Mn$_{x}$Si$_{1-x}$ measured by standard four-point probe method on lithographically defined Hall ball geometry. Insert shows the same plot in log-log scale. Insert shows the same data on log-log scale to make the insulating behavior for the $x$=0.04, 0.095, and 0.135 samples clear.}
            \label{fig:sigmaT}
\end{figure}

Figure~\ref{fig:sigmaT} shows the dc conductivity ($\sigma_\text{dc}$) vs $T$ for different $x$. The conductivity has a positive temperature coefficient and increases monotonically with the level of Mn doping, indicating semiconducting behavior and the effect of Mn doping in \textit{a}-Si. The increasing $\sigma_\text{dc}$ with increasing $T$ is also a signature of localization in a disordered electronic system. A concentration-tuned \textit{M-I} transition is visible; samples are insulating for $x$$\leq$0.135, whereas they are metallic for higher $x$. The 0.135 sample is very close to $x_{c}$ but slightly on the insulating side. This result is in good agreement with the $x_{c}$=0.137 found in Mn-implanted \textit{a}-Si samples~\cite{Yakimov1997_JETP}. 

Magneto-transport properties were measured for one sample on the insulating side ($x$=0.135) and one sample on the metallic side ($x$=0.175). A small \textit{positive} MR was found down to 2 K at $H$=7 T for both samples, 27\% and 16\% for $x$=0.135 and $x$=0.175, respectively, which can be attributed to the electron correlation effect in disordered non-magnetic electronic systems~\cite{Lee1985}. Yakimov and coworkers found a crossover from small \textit{positive} MR to small \textit{negative} MR below 2.3 K (-25\% at 1.76 K and 4.5 T) for a Mn-ion-implanted \textit{a}-Si sample ($x$=0.11) held at room temperature for 8 years~\cite{Yakimov1995}. We did not observe any negative MR in our samples. The negative MR found in Ref.~\onlinecite{Yakimov1995} was attributed to magnetic clusters.

Zero-field-cooled (ZFC) and field-cooled (FC) magnetic susceptibility $\chi(T)$ data were recorded in a dc magnetic field of 1 T. For $x$=0.04, which had the largest raw magnetic signal, ZFC and FC susceptibilities were also measured in smaller fields ($H$= 1000 and 350 Oe). All $\chi(T)$ curves were identical and showed no splitting between ZFC and FC. Measuring $\chi(T)$ at 1 T compared to 350 Oe leads to $\sim$10\% underestimation of the effective moment ($p_\text{eff}$), but it is necessary to obtain reliable magnetic signals from films ($T$ dependent) above the diamagnetic background signal (negligible $T$ dependent) for our samples, all of which show very low magnetic signals. The diamagnetic background from the SiN$_{x}$ coated Si substrate was measured for each $T$ and $H$, as was an undoped \textit{a}-Si control sample prepared using the same Si evaporation batch. The resulting net \textit{a}-Mn$_{x}$Si$_{1-x}$ response followed the Curie-Weiss (CW) law very well with a very low CW temperature ($\left|\theta\right|<$2 K). No ferromagnetic or SG states were observed in the temperature range between 2-40 K. Above 40 K, the error bars were large due to the small signal. The thermoremanent moment (TRM) was obtained by cooling the sample in a field of 7 T and then measuring the TRM on heating the sample in zero magnetic field from 1.9 K to 380 K: no significant signal was obtained at any temperature for any of the samples. These results indicate that the magnetic signal is purely paramagnetic.

\begin{figure}
    \centering
    \includegraphics[width=0.5\textwidth]{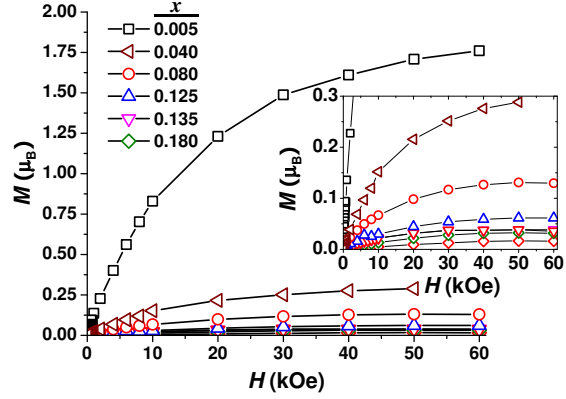}% Here is how to import EPS art
            \caption{(color online) Magnetization per Mn atom in units of $\mu_\text{B}$ vs. $H$ for different Mn concentration at 2 K, assuming all Mn are contributing (i.e. $M$=$m$/$n_\text{Mn}$ $\mu_\text{B}$ with $n_\text{Mn}$ from Table~\ref{tab:table1}). Insert shows same data on expanded scale. Lines are guides to the eye.}
            \label{fig:MvsH}
\end{figure}

\begin{figure}
    \centering
    \includegraphics[width=0.5\textwidth]{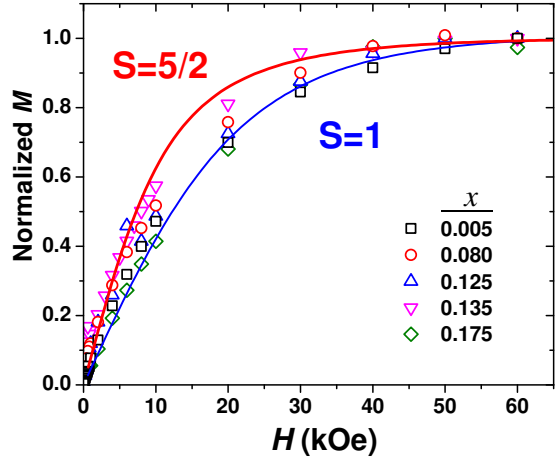}% Here is how to import EPS art
            \caption{(color online) Normalized $M$ vs $H$ for various compositions $x$. Lines are Brillouin function for $J$=$S$=5/2 and $J$=$S$=1 for comparison.}
            \label{fig:BFunction}
\end{figure}

Figure~\ref{fig:MvsH} shows $M$ (magnetization per Mn atom) vs $H$ at 2 K, assuming that all Mn atoms contribute equally to the signal. The diamagnetic background from the substrate was again subtracted. No hysteresis loop was found. $M$ decreased very fast with the Mn concentration. If it is assumed that either $J$=1 or 5/2 for $x$=0.005, then 92\% or 37\% of the Mn atoms are magnetically active, respectively. The $M$ vs $H$ data for all samples scale well to a Brillouin function for free magnetic ions, as shown in Figure~\ref{fig:BFunction}. The value of $M$ at 6 T should be within 2\% of the saturation moment for non-interacting Mn ions (whether $J$=1 or 5/2), and so its value can be used to calculated the saturation moment per Mn atom ($p_\text{sat}$) as a function of $x$.

Figure~\ref{fig:PvsX} shows the Mn concentration dependence of the $p_\text{eff}$ as taken from the CW fitting constant $A$=$n_\text{Mn}p_\text{eff}^{2}\mu_\text{B}^{2}/3k_\text{B}$ with $n_\text{Mn}$ from RBS, and of the $p_\text{sat}$ as taken from $M$ at $T$=2 K, $H$=6 T shown in Fig.~\ref{fig:MvsH}. Both $p_\text{eff}$ and $p_\text{sat}$ decrease sharply from $x$=0.005 to $x$=0.04, becoming very small and decreasing smoothly with increasing Mn concentration. 

\section{Discussion}
\begin{figure}
    \centering
    \includegraphics[width=0.5\textwidth]{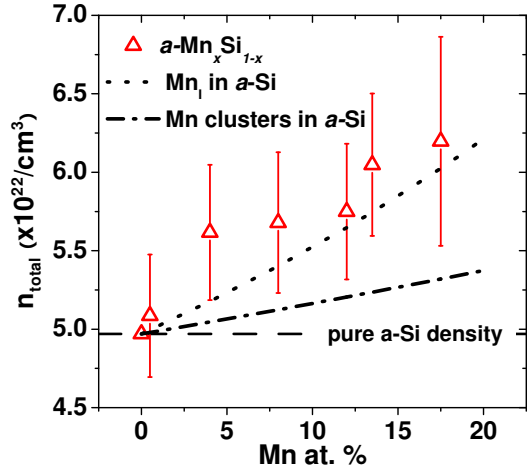}% Here is how to import EPS art
            \caption{(color online) Total atomic number density ($n_\text{total}$, obtained by dividing RBS areal density by film thickness), including both Mn and Si, as a function of $x$. For comparison, dotted line is the calculated $n_\text{total}$ assuming all doped Mn go in interstitially without affecting the \textit{a}-Si structure; dash-dot line is the calculated $n_\text{total}$ assuming the same fraction of Mn are totally phase segregated into metal clusters in \textit{a-}Si [number density of pure \textit{a-}Si (4.97$\times$10$^{22}$/cm$^{3}$) and Mn metal (7.96$\times$10$^{22}$/cm$^{3}$) are used].}
            \label{fig:Ndensity}
\end{figure}

As shown in Fig.~\ref{fig:Ndensity}, the total atomic number density $n_\text{total}$ increases with Mn doping. This is comparable to, or slightly higher than, the calculated $n_\text{total}$ (also plotted in Fig.~\ref{fig:Ndensity}) assuming all Mn are at interstitial sites without changing the \textit{a}-Si matrix density, suggesting that Mn atoms act as interstitial dopants in \textit{a}-Mn$_{x}$Si$_{1-x}$ (therefore are at sites with high Si coordination number and low symmetry). Mn$_\text{Si}$ on the other hand would keep $n_\text{total}$ constant. The charge and spin states of Mn strongly depend on the local environment. Preliminary ESR results on our films show $g=$2.01$\pm$0.03 for Mn, indicating Mn is in a pure spin state. Based on the LW model proposed for Mn in Si, there are two possible interstitial Mn sites with quenched orbital moment: one is Mn$_\text{I}^{2+}$ (3$d^{5}$, $J$=$S$=5/2) and the other is Mn$_\text{I}^{-}$ (3$d^{8}$, $J$=$S$=1). 

\begin{figure}
    \centering
    \includegraphics[width=0.5\textwidth]{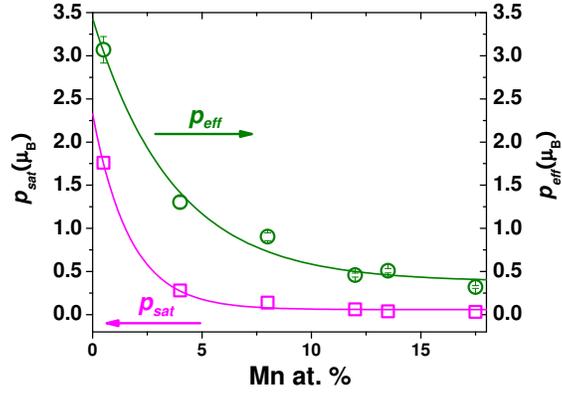}% Here is how to import EPS art
            \caption{(color online) $p_\text{eff}$ (extrapolated from $A=n_\text{Mn}p_\text{eff}^{2}\mu_\text{B}^{2}/3k_\text{B}$ 
            obtained by the CW fitting) and $p_\text{sat}$ 
            (calculated from $M_\text{sat}$=$n_\text{Mn}p_\text{sat}$ with $M_\text{sat}$ 
            taken from $M$ at $T=$2 K, $H=$6 T shown in Fig.~\ref{fig:MvsH}) 
            for different Mn concentrations based on $n_\text{Mn}$ 
            from RBS as shown in Table~\ref{tab:table1}. Lines are guides to the eye.}
            \label{fig:PvsX}
\end{figure}

The small magnetization (both $p_\text{eff}$ and $p_\text{sat}$) of all samples suggests that only a very small fraction of Mn atoms are magnetically active. Based on the saturation moment, we can obtain the fraction of magnetically active Mn ions (denoted as $c_{m}$) assuming $J$=$S$=1 or 5/2. For a pure paramagnetic state, the same magnetic centers contribute to $\chi(T)$. The atomic moment ($gJ$) and the effective moment ($g\sqrt{J(J+1)}$) are connected for a specific $J$ value, with $g$=2 from ESR data. We re-calculate the new effective moment (now denoted as $p^{m}_\text{eff}$) based on $c_{m}n_\text{Mn}$ (instead of $n_\text{Mn}$ in Fig.~\ref{fig:PvsX}) and the measured CW constant $A$. For $J$=$S$=1, the resulting $p^{m}_\text{eff}$ values are more than 20\% \textit{greater} than the theoretical value of 2.83 $\mu_\text{B}$. $\chi$ was determined with a 1 T field, which leads to smaller $\chi$ than determined at lower field, thus underestimating of $p^{m}_\text{eff}$. This indicates that $J$=$S$=1 is not the correct spin state. For $J$=$S$=5/2, the resulting $p^{m}_\text{eff}$ values are $\sim$10\% \textit{smaller} than the theoretical value of 5.92 $\mu_\text{B}$. We can check the underestimation of $p^{m}_\text{eff}$ by using the Brillouin function for $J$=$S$=5/2 moment at 2 K. Since $p_\text{eff}$$\propto$$\sqrt{A}$$\propto$$\sqrt{\chi}$, the fractional reduction $\Delta p_\text{eff}/p_\text{eff}$$\propto$$\frac{1}{2}\Delta\chi / \chi$. The calculated $\Delta\chi / \chi$ from using 1 T instead of the low-field limit is $\sim$-20\%, which gives $\Delta p_\text{eff}/p_\text{eff}\sim$-10\%. Taking this -10\% deviation into account, $p^{m}_\text{eff}$ and $p_\text{sat}$ are in excellent agreement for $J$=$S$=5/2. In Table~\ref{tab:table1}, we summarize all of the fitting parameters as well as the estimated $c_{m}$ and $p^{m}_\text{eff}$. The agreement between the atomic moment and the effective moment for all $x$ values strongly suggests that Mn$_\text{I}^{2+}$ (3$d^{5}$, $J$=$S$=5/2) are the magnetically active sites in these films, but only account for a small fraction of the total doping (e.g. $c_{m}$=0.6\% for $x$=0.175). 

The decreasing moment seen in Fig.~\ref{fig:PvsX} can be explained by as the decrease of $c_{m}$ with $x$. $c_{m}$ are 35.2\% and 5.6\% for the $x$=0.005 and 0.04 films, respectively, and become less than 1\% when $x\geq$0.12 (as shown in Table~\ref{tab:table1}). The rest of the Mn atoms are in a non-magnetic state and contribute to the transport properties only (as shown by the increase of $\sigma_\text{dc}$ with $x$). 

These two types of Mn states found in \textit{a}-Mn$_{x}$Si$_{1-x}$ are in sharp contrast to the situation for \textit{a}-Gd$_{x}$Si$_{1-x}$, where all Gd atoms are in Gd$^{3+}$ states, contributing 3$e^{-}$ as well as 7 $\mu_\text{B}$ moment (due to the half-filled $f$ shell, $J$=$S$=7/2) per Gd. The dual role of Gd is the key to its SG ground state as well as the enormous negative MR in the magnetically doped amorphous semiconductor studied previously ~\cite{Zeng2007PRB}. In \textit{a}-Mn$_{x}$Si$_{1-x}$, only a tiny percentage of the Mn sites are magnetically active and therefore far separated, and thus no magnetic interaction is developed as in \textit{a}-Gd$_{x}$Si$_{1-x}$. The majority of Mn atoms are in a non-magnetic state (which is not predicted by the LW model), despite contributing to transport. The small positive MR is consistent with this result, as in other non-magnetic disordered systems~\cite{Lee1985}. We do not observe any negative MR in our co-deposited \textit{a}-Mn$_{x}$Si$_{1-x}$ films even for samples on the insulating side, as reported previously for Mn-implanted samples~\cite{Yakimov1995}. 

Preliminary X-ray absorption spectroscopy (XAS) data on the Mn $L_{2}$ and $L_{3}$ edges for all samples with $x$$\geq$0.04 show two peaks with very broad features, quite different from the spectra seen for Mn materials with localized $d$ electrons which show distinct multiplets. These broad peaks are evidence that the majority of Mn form an impurity band in the \textit{a}-Si matrix. States in this band are localized due to disorder when $x$$\leq$$x_{c}$, and delocalized when $x$$>$$x_{c}$. Preliminary Hall measurements using lithographically defined Hall patterns suggest an electron-like carrier type for the metallic samples. 

\begin{figure}
    \centering
    \includegraphics[width=0.5\textwidth]{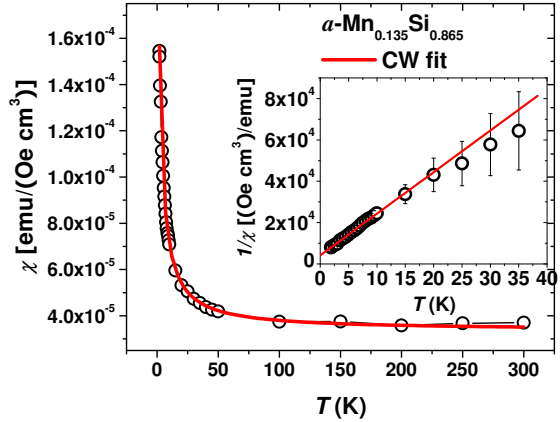}% Here is how to import EPS art
            \caption{(color online) ZFC $\chi(T)$ data for \textit{a}-Mn$_{0.135}$Si$_{0.865}$, measured on heating in an 1 T dc magnetic field after cooling the sample in zero field. FC $\chi(T)$ is identical. Other concentrations fit similarly, but with $p_\text{eff}$ dependent on $x$. Solid lines are the fit to CW law.}
            \label{fig:ChiT}
\end{figure}

How would the amorphous matrix affect the local moment of the Mn$_\text{I}$ sites in \textit{a}-Si compared to the \textit{c-}Si case? Mn$_\text{I}$ sites in \textit{c-}Si do not form covalent bonds with Si and are always predicted to have local moments. This should be the same case in \textit{a}-Si. However, if any local moment exists in our \textit{a}-Mn$_{x}$Si$_{1-x}$ films, for the wide concentration and the temperature range measured, one would not expect a purely paramagnetic response with small $\theta$ values and no sign of any magnetic interaction. It is possible but unlikely that local moments exist but are completely canceled because of Mn-Mn AFM interactions, since AFM ordering is not robust to disorder and should lead to a magnetic freezing state, such as a SG phase (with $T_{f}$) or a clustered spin glass (with blocking temperature, $T_{B}$), showing magnetic hysteresis (differences between FC and ZFC states) and TRM. $T_{f}$ or $T_{B}$ should increase with the magnetic concentration, but no magnetic freezing was observed for $x$ up to 0.175. It is possible that $T_{f}$ and $T_{B}$ are so high that samples are already in a frozen state at 40 K and thus only express a very small moment. However, known concentrated SG \textit{a}-MnSi (1:1 stoichiometry) has a $T_{f}$ only at 22 K~\cite{Hauser1979}, while the CW behavior of our samples is reliably measured up to 40 K, as shown in Fig.~\ref{fig:ChiT} (above 40 K, the signal is too small with large error bars). $T_{f}$ of \textit{a}-MnSi is smaller but comparable to the magnetic ordering temperature ($T_{C}$=30 K) of its crystalline counterpart, compound MnSi, which has a helical spin structure with $\sim$18 nm wavelength. Above $T_{f}$ or $T_{C}$, in the paramagnetic states, they have comparable $p_\text{eff}$, which are 2.2 and 2.6 $\mu_\text{B}$ for the compound MnSi and \textit{a}-MnSi respectively~\cite{Cochrane1979}; both of these states would give significantly larger $\chi$ than observed at 40 K. 

If there were non-interacting Mn complex clusters in \textit{a-}Si (invisible in HR-XTEM), such as a Mn$_\text{I}$-Mn$_\text{I}$ dimer instead of interacting Mn$_\text{I}$ ions, these could be strongly AFM-coupled. Such complexes have not been experimentally observed to the best of our knowledge, and are calculated to be highly unstable energetically~\cite{Bernardini2004}. Furthermore, only the neutral charge state favors AFM coupling in Si, which should therefore not contribute to the transport. Mn$^{0}_{4}$~\cite{Ludwig1960} and [Mn$^{0}_\text{3I}$-Mn$^{-}_\text{I}$]~\cite{Kreissl1994} clusters have large $S$ values, thus not the case here. Existence of metallic Mn or Mn-rich clusters of larger scale are not supported by the HR-XTEM results (shown in Fig.~\ref{fig:HRXTEM}) and the atomic density analysis (shown in Fig.~\ref{fig:Ndensity}). Moreover, ferromagnetic clusters should result in an enhanced $p_\text{eff}$, opposite to what we have seen. All these strongly indicate the majority Mn atoms enter into a non-magnetic doping environment in \textit{a-}Si, forming even at $x$ as low as 0.005 ($\sim$64.8\%). 

The totally quenched Mn moment in \textit{a}-Si is intriguing and needs more understanding. 
In our two-state scenario, we have argued that the small magnetic signal is due to a small fraction 
of Mn$_\text{I}^{2+}$ sites according to the LW model, while the majority of Mn atoms are in a 
totally non-magnetic state. Based on the comparable (high) atomic number density 
and the CRN model for \textit{a}-Si, the local site symmetry should be only slightly perturbed for these 
Mn$_\text{I}$ when $x$ is small, and thus the LW model is still valid at least for low $x$. If adding more Mn enhances 
the crystal field splitting such that it overrules the LW model and Hund's rule, a zero moment state 
could arise (Mn$_\text{I}^{+}$ with filled $t_{2}$ levels, or Mn$_\text{I}^{3+}$ with filled $e$ levels, 
dependent on which is lower), although such charge states should have multiplets at the $L$ edges in 
XAS measurement due to unfilled $d$ levels. Another scenario is itinerant magnetism with very small 
moment, as found in some Mn silicides, such as the crystalline compound Mn$_{4}$Si$_{7}$, where 
$p_\text{eff}$ (from CW fit for $T$$>$40 K, $\theta$=43 K) and $p_\text{sat}$ (from saturation $M$ at 2 K) were found to be 0.365 and 0.012 $\mu_\text{B}$ respectively~\cite{Gottlieb2003}. 
The space group of the tetragonal Mn$_{4}$Si$_{7}$ is $P\bar{4}c2$. The symmetry at the Mn sites 
in this compound is low and may be a better representation of the local environment of Mn in 
\textit{a-}Mn$_{x}$Si$_{1-x}$. The large $p_\text{eff}$/$p_\text{sat}$ in our samples 
(as shown in Table~\ref{tab:table1}) may also suggest the existence of itinerant moments~\cite{Rhodes1963}. The existence of itinerant moments in insulating samples is compatible with an Anderson localization model for the effect of strong disorder. 

It would be useful to evaluate how disorder affects the charge and spin states of Mn using first-principles calculations. If the analog of the local environment between \textit{a-}Si and \textit{c-}Si is valid, and since the local moment is largely determined by the local environment, then Mn would be expected to be in an interstitial site with $J$=$S$=5/2. However, our results indicate almost no moment for Mn, suggesting a new state for Mn in \textit{a}-Si, which is not predicted by any existing model for Mn in \textit{c-}Si. 

\section{Conclusion}
Magnetic and electrical transport properties were measured for $e$-beam co-deposited homogeneous \textit{a}-Mn$_{x}$Si$_{1-x}$ with $x$ from 0.005 to 0.175. A small fraction of Mn in a Mn$^{2+}_\text{I}$ configuration account for the small detected paramagnetic signal. The majority of Mn atoms are, however, in a totally non-magnetic state, which is not observed in the crystalline counterpart or predicted by any existing model for transition metal impurities in Si. This explains why MR for \textit{a-}Mn$_{x}$Si$_{1-x}$ is positive and small, with typical values for disordered non-magnetic electronic systems, unlike \textit{a-}Gd$_{x}$Si$_{1-x}$, where an enormous negative MR was measured and explained in terms of interplay between large Gd local moments and charge carriers. The non-magnetic state is most likely due to the formation of an impurity band, subject to localization effects, although a tightly bound dimer state cannot be ruled out.

\begin{acknowledgments}
We thank S. Lofland for ESR data, E. Arenholz, E. Cruz and C. Piamonteze for assistance with XAS experiment, N. Spaldin, A. J. Freeman, Hua Wu and M. Scheffler  for useful discussions and D. R. Queen for assistance. We also acknowledge use of facilities in the LeRoy Eyring Center for Solid State Science at Arizona State University. This research was supported by NSF DMR-0505524 and 0203907.
\end{acknowledgments}

\newpage %Just because of unusual number of tables stacked at end

%\bibliography{aMnSi}% Produces the bibliography via BibTeX.

\newpage

\begin{table*}
\caption{\label{tab:table1}Sample composition, Curie-Weiss parameters and magnetic moment analysis for \textit{a-}Mn$_{x}$Si$_{1-x}$ based on two different methods of treating the fraction of magnetically active Mn.}
\begin{ruledtabular}
\begin{tabular}{ccccccdc}
$x$\footnotemark[1] & $n_\text{Mn}$\footnotemark[1] & $\theta$\footnotemark[2] & $A$\footnotemark[2] & $p_\text{eff}$\footnotemark[3] & \mbox{$p_\text{eff}$/$p_\text{sat}$}\footnotemark[3] & \mbox{$c_{m}$}\footnotemark[4] & $p^{m}_\text{eff}$\footnotemark[5] \\
&($\times$10$^{21}$ atoms/cm$^3$) &  (K) & ($\times$10$^{-4}$ emu/cm$^{3}\cdot$Oe$\cdot$K) & ($\mu_\text{B}$) &  & (\%) &  ($\mu_\text{B}$) \\
\hline
0.005& 0.27 & -1.63 & 5.2 & 3.1 & 1.74  & 35.2 & 5.2 \\
0.04 & 2.04 & -0.45 & 7.2 & 1.3 & 4.39  & 5.6 & 5.5 \\
0.08 & 4.54 & -1.71 & 7.7 & 0.9 & 6.57  & 2.8 & 5.4 \\
0.12 & 6.86 & -0.19 & 3.0 & 0.5 & 7.46  & 1.2 & 4.1 \\
0.135& 8.20 & -1.69 & 4.4 & 0.5 & 13.2 & 0.8 & 5.8 \\
0.175& 10.8 & -1.44 & 2.3 & 0.3 & 9.99  & 0.6 & 4.0 \\
\end{tabular}
\end{ruledtabular}
\footnotetext[1] {From RBS and AFM measurements.} 
\footnotetext[2] {$\theta$ and $A$ from CW fit: $\chi=A/(T-\theta)+b$; $b$ is a small temperature-independent constant due to combined core, and magnetometer background contributions}
\footnotetext[3] {$p_\text{eff}$ per Mn obtained from $A=n_\text{Mn}p^{2}_\text{eff}\mu^{2}_\text{B}/3k_\text{B}$; $p_\text{sat}$ per Mn obtained from saturation moment at 2 K and 6 T, both assuming all $n_\text{Mn}$ contribute equally, with $n_\text{Mn}$ from RBS.} 
\footnotetext[4] {$c_{m}$ is the magnetically active Mn concentration obtained from saturation magnetization, assuming each active Mn has $J$=$S$=5/2.}
\footnotetext[5] {$p^{m}_\text{eff}$ is the calculated effective moment per active Mn atom based on $c_{m}$, each with $J$=$S$=5/2.} 
\end{table*}

\newpage

%\begin{figure}
%    \centering
%    \includegraphics[width=0.5\textwidth]{sigmatime}% Here is how to import EPS art
%            \caption{Room temperature conductivity vs. time exposed in air after taken out from the vacuum deposition chamber.}
%            \label{fig:sigmatime}
%\end{figure}

\end{document}